\documentclass[aps,onecolumn,pra,showpacs,amsmath,amssymb,showkeys,10pt]{revtex4-2}
\usepackage{graphicx}
\usepackage{epstopdf}
\usepackage{diagbox}
\usepackage{multirow}
\pdfoutput=1
\usepackage{braket}
\usepackage{hyperref}
\usepackage{longtable}

\begin{document}

	\title{Single-Photon Motion in a Two-Dimensional Plane: Confinement and Boundary Escape}
	
	\author{Hui-hui Miao}
	\email[Correspondence to: Vorobyovy Gory 1, Moscow, 119991, Russia. E-mail address: ]{hhmiao@cs.msu.ru}
	\affiliation{Faculty of Computational Mathematics and Cybernetics, Lomonosov Moscow State University, Vorobyovy Gory 1, Moscow, Russia}

	\date{\today}

	\begin{abstract}
	This paper investigates the motion of a single photon in a two-dimensional plane under closed and open boundary conditions. We employ two methods to construct the Hilbert space: Method A, based on the standard second-quantization formalism, and Method B, based on a non-standard approach. By eliminating redundant quantum states, we obtain a reduced Hilbert space with significantly lower dimensionality, thereby improving the efficiency of numerical simulations. In a closed system, the two methods are equivalent, and their unitary evolution results are identical. The probability distribution diffuses outward from the center and exhibits a significant rebound after reaching the boundary. In an open system, Method B, by incorporating more dissipation channels, provides a more accurate description of the photon escape process at the boundary. The probability curves obtained from the two methods completely overlap before reaching the boundary. After the boundary is reached, a slight difference appears, but this difference does not amplify with evolution and tends to converge in the later stage. Method B yields a slightly higher dissipative-state probability, indicating that the photon escapes faster. Visualization of the two-dimensional probability distribution shows that the three scenarios (closed system, open system with Method A, and open system with Method B) exhibit identical probability distributions before reaching the boundary. After the boundary is reached, the open systems exhibit significant probability loss, which increases rapidly with evolution. The probability distribution patterns of the two open systems are highly similar, exhibiting synchronized evolution. By comparing two Hilbert space construction methods, this paper reveals the motion of a single photon in a two-dimensional plane, providing a theoretical foundation for the study of photon transport and dissipation control in complex quantum networks.
	\end{abstract}


	\maketitle

	\section{Introduction}
	\label{sec:Intro}
	
	Cavity quantum electrodynamics (QED) models, as an ideal platform for studying light--matter interactions, have been extensively investigated. In recent years, research on these models has deepened, covering multiple cutting-edge directions such as quantum phase transitions \cite{Prasad2018, Wei2021}, quantum gates \cite{OzhigovYI2020, Dull2021}, quantum correlations \cite{Miao2024, MiaoLi2025, Miao2025}, dark states \cite{Lee1999, Andre2002, Poltl2012, Tanamoto2012, Hansom2014, Kozyrev2018, Ozhigov2020}, and other topics \cite{Guo2019, Victorova2020, Kulagin2022, Afanasyev2022, Pluzhnikov2022, MiaoOzhigov2024, LiMiao2024}. Typical cavity QED models include the Jaynes-Cummings model \cite{Jaynes1963} and the Tavis-Cummings model \cite{Tavis1968}, as well as their generalizations \cite{Angelakis2007}. These models typically consist of optical microcavities and artificial atoms, where the coupling between the cavity field and the atoms enables the simulation of a wide range of quantum phenomena. However, the modeling and simulation of quantum systems often lead to an exponential increase in complexity as the number of degrees of freedom grows. This computational challenge, arising from increased dimensionality, was first identified by R.E. Bellman and termed the "curse of dimensionality" \cite{Bellman1957, Bellman1961}. In our previous work, we have attempted to address some of the computational issues caused by the curse of dimensionality using numerical methods \cite{Miao2023, You2023, Chen2023, MiaoOzhigov2024, LiMiao2024}, such as obtaining a reduced Hilbert space by eliminating redundant quantum states \cite{Miao2023}.

	This paper focuses on the motion of a single photon in a two-dimensional plane, systematically studying its quantum dynamics under closed and open boundary conditions. We employ two methods to construct the Hilbert space of the system: Method A, based on the standard second-quantization formalism, and Method B, based on a non-standard approach. By eliminating redundant quantum states that do not participate in the dynamics, we obtain a reduced Hilbert space with significantly lower dimensionality, thereby improving the efficiency of numerical simulations. Based on this, we compare the evolution results obtained from the two methods in closed and open systems, analyzing the diffusion of the probability distribution, the boundary bounce, and the effect of dissipation channels on the photon escape process. The remainder of this paper is organized as follows. Sec. \ref{sec:TheoModel} introduces the theoretical model and the two methods for constructing the Hilbert space. Sec. \ref{sec:Method} describes the numerical method. Sec. \ref{sec:Results} presents the numerical simulation results, including the probability evolution in closed and open systems, a comparative analysis of the two methods, and a visualization of the two-dimensional spatial distribution. Sec. \ref{sec:Conclusion} summarizes the paper and provides an outlook on future research directions.
	
	\section{Theoretical model}
	\label{sec:TheoModel}
	
	\begin{figure}
		\centering
		\includegraphics[width=1.\textwidth]{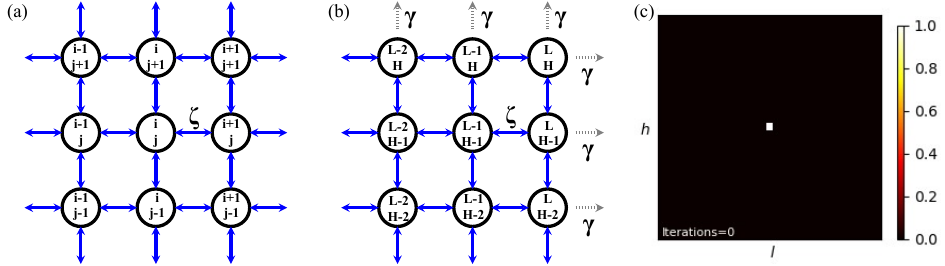} 
		\caption{(online color) {\it Schematic diagram of single-photon motion in a two-dimensional plane.} Panel (a) shows the coupling scheme between optical cavities. All optical cavities are arranged in a rectangular (or square) grid on the plane. Each cavity is connected to its four nearest neighbors via optical fibers along the horizontal and vertical directions. The photon tunneling strength between different cavities is $\zeta$. We use $l$ and $h$ to denote the horizontal and vertical coordinates of the cavity, respectively. The total numbers of cavities along the horizontal and vertical directions are denoted as $l_{\max}=L$ and $h_{\max}=H$, respectively. Under open boundary conditions, dissipation channels exist on all four boundaries of this two-dimensional rectangular plane, with a dissipation strength $\gamma$. Panel (b) illustrates this dissipation channel, using the upper-right corner as an example. Panel (c) shows the initial state of the system: the single photon is initially located at the exact center of the two-dimensional plane.}
		\label{fig:TheoModel}
	\end{figure}
	
	This section introduces the Jaynes--Cummings--Hubbard model (JCHM) \cite{Jaynes1963, Angelakis2007} to describe the motion of a single photon in a two-dimensional plane composed of numerous optical cavities. The arrangement of the optical cavities is shown in Fig. \ref{fig:TheoModel}(a), and the cavities are coupled together via optical fibers. The system contains only one free photon, which can transition between different cavities via the tunneling coefficient $\zeta$. We consider two boundary cases:
	\begin{itemize}
		\item Closed boundary: the photon cannot leak out of the system, and the corresponding quantum dynamics are unitary.
		\item Open boundary: the photon can escape from the boundary to the external environment, resulting in dissipative dynamics of the system (see Fig. \ref{fig:TheoModel}(b)).
	\end{itemize}
	
	First, we employ the second-quantization formalism \cite{Dirac1927, Fock1932} to construct the Hilbert space. The total Hilbert space then takes the form:
	\begin{equation}
		\label{eq:MethodA}
		|\Psi\rangle_{\mathcal{C}}=\bigotimes_{l,h}|p_{l,h}\rangle_{l,h},
	\end{equation}
	where $l$ and $h$ are the coordinates of the cavity along the horizontal and vertical directions, respectively, with $0\leq l\leq L$ and $0\leq h\leq H$. Here, $L$ and $H$ denote the total numbers of cavities in the horizontal and vertical directions, respectively. We define the variable $p_{l,h}\in\{0,1\}$ to indicate whether the cavity at coordinates $(l,h)$ is occupied by a photon: $p_{l,h}=0$ means the cavity is empty, and $p_{l,h}=1$ means the cavity is occupied by a single photon. The Hilbert space dimension of the entire system is $2^{L\times H}$, where $L\times H$ is the total number of optical cavities on the plane. Since our model contains only one photon, we do not need to use the full Hilbert space $\mathcal{C}$ when constructing the Hamiltonian. By eliminating the quantum states that do not participate in the dynamics, we obtain a reduced Hilbert space $\mathcal{C}'$, which takes the form:
	\begin{equation}
		\label{eq:MethodASample}
		|\Psi\rangle_{\mathcal{C}'}=\bigotimes_{\substack{l,h\\\sum_{l,h}p_{l,h}\leq 1}}|p_{l,h}\rangle_{l,h}.
	\end{equation}
	The dimension of the reduced Hilbert space $\mathcal{C}'$ in Eq. \eqref{eq:MethodASample} is $L\times H$ for the closed system and $L\times H+1$ for the open system.
	
	In addition to the standard second-quantization formalism, we propose an alternative method for constructing the Hilbert spaces:
	\begin{equation}
		\label{eq:MethodB}
		|\Psi\rangle_{\mathcal{C}}=|l\rangle|h\rangle|p\rangle|m_{\leftarrow}\rangle|m_{\rightarrow}\rangle|m_{\uparrow}\rangle|m_{\downarrow}\rangle,
	\end{equation}
	where $l$ and $h$ also denote the horizontal and vertical coordinates of the cavity where the photon was last located. The variable $p\in{0,1}$ indicates whether the photon is still in the system: $p=1$ means the cavity at $(l,h)$ is occupied, and $p=0$ means the photon has escaped to the external environment. The variables $m_{\leftarrow}, m_{\rightarrow}, m_{\uparrow}, m_{\downarrow}\in{0,1}$ indicate the escape direction: $m_{\leftarrow}=1$ ($m_{\rightarrow}=1$) means the photon escapes horizontally to the left (right), and $m_{\uparrow}=1$ ($m_{\downarrow}=1$) means it escapes vertically upward (downward). Otherwise, these variables are $0$. The parameters $p$, $m_{\leftarrow}$, $m_{\rightarrow}$, $m_{\uparrow}$, and $m_{\downarrow}$ always satisfy the constraint:
	\begin{equation}
		\label{eq:Condition}
		p+m_{\leftarrow}+m_{\rightarrow}+m_{\uparrow}+m_{\downarrow}=1.
	\end{equation}
Accordingly, the Hilbert space dimension of the system is $5\times L\times H$. Eq. \eqref{eq:MethodB} accounts for the single-photon case. Therefore, for sufficiently large $L$ and $H$, the dimension of $\mathcal{C}$ in Eq. \eqref{eq:MethodB} is much smaller than that in Eq. \eqref{eq:MethodA}, which is $2^{L \times H}$.
	
	The core difference between Eq. \eqref{eq:MethodA} and Eq. \eqref{eq:MethodB} is as follows. In Eq. \eqref{eq:MethodA}, each optical cavity has one dissipation channel leading to the same target state $\bigotimes_{l,h}|0\rangle_{l,h}$. This means that, regardless of which cavity the photon occupies, the dissipation always leads to the same quantum state. In contrast, in Eq. \eqref{eq:MethodB}, each cavity has four dissipation channels (left, right, up, and down), each leading to a distinct quantum state. The latter thus represents a more realistic physical model.
	
	 The Hilbert space obtained from Eq. \eqref{eq:MethodB} can be further reduced by removing redundant states. Since the photon can only escape to the external environment through the four boundaries, $p$ can only be equal to $1$ when the photon is not at the boundary, i.e., when $0<l<L$ and $0<h<H$. In this case, $m_{\leftarrow}=m_{\rightarrow}=m_{\uparrow}=m_{\downarrow}=0$. When the photon is located at the boundary, the following four cases arise:
	\begin{itemize}
		\item When $l=0$, the photon is on the left boundary. Through dissipation, $p$ changes from $1$ to $0$, while $m_{\leftarrow}$ changes from $0$ to $1$, indicating that the photon escapes horizontally to the left.
		\item When $l=L$, the photon is on the right boundary. Through dissipation, $p$ changes from $1$ to $0$, while $m_{\rightarrow}$ changes from $0$ to $1$, indicating that the photon escapes horizontally to the right.
		\item When $h=0$, the photon is on the bottom boundary. Through dissipation, $p$ changes from $1$ to $0$, while $m_{\downarrow}$ changes from $0$ to $1$, indicating that the photon escapes vertically downward.
		\item When $h=H$, the photon is on the top boundary. Through dissipation, $p$ changes from $1$ to $0$, while $m_{\uparrow}$ changes from $0$ to $1$, indicating that the photon escapes vertically upward.
	\end{itemize}
	Additionally, when the photon is located at the intersection of two boundaries, it has two dissipation channels simultaneously. The dimension of the reduced Hilbert space $\mathcal{C}'$ is therefore $L\times H$ for the closed system and $L\times H+2\times(L+H)$ for the open system.
	
	For convenience, we refer to the method of constructing the Hilbert space via the standard second-quantization formalism as Method A, and the non-standard method as Method B. A comparison of the two methods is shown in Tab. \ref{Tab:ComparisonMethods}, which demonstrates that for sufficiently large $L$ and $H$, $\dim{\mathcal{C}'}\ll \dim{\mathcal{C}}$ for both the closed and open systems. This redundancy removal significantly reduces memory usage and improves computational efficiency in numerical simulations.
	
	\begin{table}[!htpb]
        \centering
		\begin{tabular}{|c|c|c|c|}
			\hline
			\multirow{2}{*}{} & \multirow{2}{*}{$\dim(\mathcal{C})$} & \multicolumn{2}{|c|}{$\dim(\mathcal{C}')$} \\
			\cline{3-4}
			& & Closed system & Open system \\
			\hline
			Method A & $2^{L\times H}$ & $L\times H$ & $L\times H+1$ \\
			\hline
			Method B & $5\times L\times H$ & $L\times H$ & $L\times H+2\times(L+H)$ \\
			\hline
		\end{tabular}
		\caption{{\it Comparison of two methods for constructing the Hilbert spaces: Method A and Method B.}}	
		\label{Tab:ComparisonMethods}
	\end{table}
	
	The Hamiltonian of the system is given by the energy operator under the rotating-wave approximation \cite{wu2007}, with $\zeta\ll\hbar\omega$:
	\begin{equation}
		\label{eq:Hamiltonian}
		\hat{H}=\sum_{l,h}^{L,H}\hbar\omega a_{l,h}^{\dag}a_{l,h}+\zeta\sum_{\substack{l_1>l_2,h_1>h_2\\l_1-l_2+h_1-h_2=1}}^{L,H}\left(a_{l_1,h_1}^{\dag}a_{l_2,h_2}+a_{l_1,h_1}a_{l_2,h_2}^{\dag}\right)
	\end{equation}
	where $\hbar$ is the reduced Planck constant, $\omega$ is the photonic mode frequency, $a^{\dag}$ is the photon creation operator, $a$ is the photon annihilation operator, and $\zeta$ is the tunneling strength.
	
	We now consider the initial state shown in Fig. \ref{fig:TheoModel}(c), with the photon located at the center of the plane. In our model, we set $L=H=31$, so the grid contains $31\times31=961$ cavities, with the four corners located at $(0,0)$, $(30,0)$, $(0,30)$, and $(30,30)$.
	
	\section{Numerical method}
	\label{sec:Method}
	
	\begin{figure}
		\centering
		\includegraphics[width=.6\textwidth]{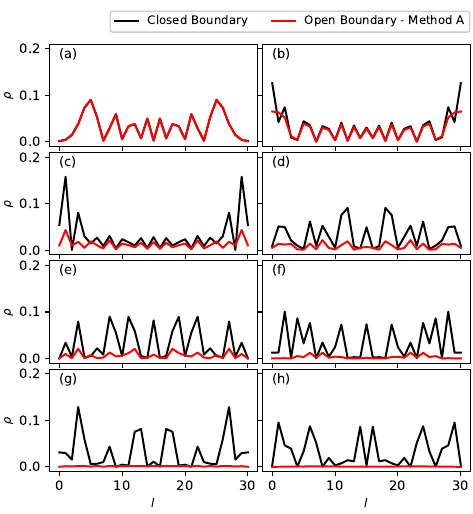} 
		\caption{(online color) {\it Comparison of evolution: unitary in the closed system vs non-unitary in the open system.} Since the photon is initially located at the center of the square plane, the probability distribution remains centrally symmetric throughout the subsequent evolution. To facilitate the analysis, we sum the probabilities over all cavity sites with the same horizontal coordinate $l$, i.e., we integrate along the vertical direction. Black and red curves represent the results of unitary and non-unitary evolution, respectively. Panels (a)--(h) show the probability distribution curves at iteration steps 300, 400, 500, 750, 1000, 2000, 5000, and 10000, respectively.}
		\label{fig:ComparisonClosedOpen}
	\end{figure}
	
	The quantum master equation (QME) for the density operator $\rho$ of the system, under the Markovian approximation, takes the form:
	\begin{equation}
		\label{eq:QME}
		i\hbar\dot{\rho}=\left[\hat{H},\rho\right]+iL\left(\rho\right)
	\end{equation}
	where $\hat{H}$ is the Hamiltonian. And the Lindblad term $L\left(\rho\right)$ accounts for the dissipation process. Thus, an approximate solution $\rho(t)$ to Eq. \eqref{eq:QME} can be obtained as follows:
	\begin{equation}
		\label{eq:UnitaryPart}
		\rho\left(t+dt\right)=\exp\left(-\frac{i}{\hbar}\hat{H}dt\right)\rho\left(t\right)\exp\left(\frac{i}{\hbar}\hat{H}dt\right)
	\end{equation}
	where $dt$ is the time step. In the second step, we advance the solution of Eq. \eqref{eq:QME} by one time step with the commutator omitted:
	\begin{equation}
		\label{eq:Solution}
		\rho\left(t+dt\right)=\tilde{\rho}\left(t+dt\right)+\frac{1}{\hbar}L\left(\tilde{\rho}(t+dt)\right)dt.
	\end{equation}
	
	\section{Results}
	\label{sec:Results}
	
	\begin{figure}
		\centering
		\includegraphics[width=.6\textwidth]{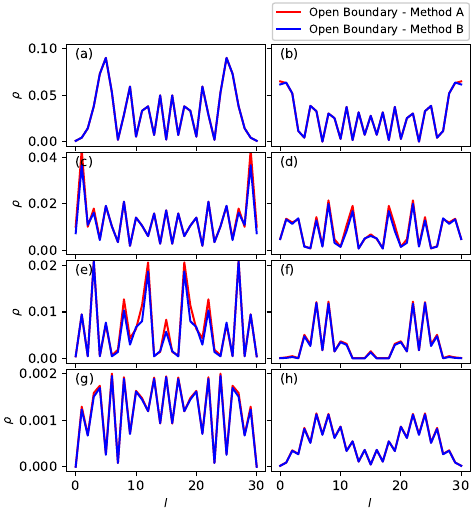} 
		\caption{(online color) {\it Probability distributions for non-unitary evolution in an open system: Method A vs Method B.} Similarly, we sum the probabilities over all cavity sites with the same horizontal coordinate $l$. Red and blue curves represent the results of non-unitary evolution obtained from Method A and Method B, respectively. Panels (a)--(h) show the probability distribution curves at iteration steps 300, 400, 500, 750, 1000, 2000, 5000, and 10000, respectively. Unlike in Fig. \ref{fig:ComparisonClosedOpen}, the vertical axis range for each row of panels has been adjusted appropriately to facilitate comparison between the red and blue curves.}
		\label{fig:ComparisonOpenOpen}
	\end{figure}
	
	In simulations: $\hbar=1$, $\omega^{\uparrow}=10^8$, and $\zeta=10^6$.
	
	Fig. \ref{fig:ComparisonClosedOpen} compares the probability distributions under closed and open boundary conditions. For the closed system, since the Hilbert spaces constructed via Method A and Method B are equivalent, we only use Method A for the unitary evolution calculations. For the open system, however, the Hilbert spaces corresponding to Method A and Method B are not equivalent. We therefore first obtain the non-unitary evolution results using Method A for comparison with the previous unitary evolution (The comparison of non-unitary evolution under the two methods will be shown in Fig. \ref{fig:ComparisonOpenOpen}). Under the centrosymmetric condition, the probability distribution along the horizontal direction remains symmetric. The probability of finding the photon gradually diffuses from the center toward both sides. As the diffusion range increases, the peak of the probability curve gradually decreases, indicating that the photon tends to become uniformly distributed across the cavities over time. It is worth noting that in the closed system, when the probability distribution reaches the boundary, the curve exhibits a significant rebound. In contrast, in the open system, this rebound is significantly weaker, indicating that the dissipation channels allow the photon to escape quickly. At 10,000 iterations (panel (h)), the red curve representing non-unitary evolution ceases to fluctuate, indicating that the photon has completely escaped from the system.
	
	\begin{figure}
		\centering
		\includegraphics[width=1.\textwidth]{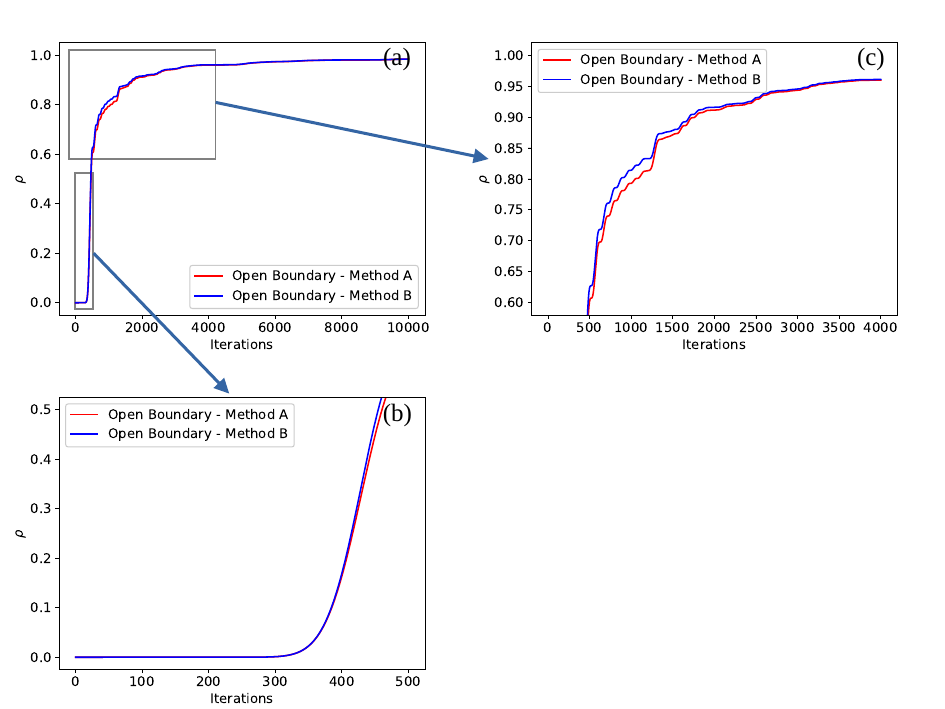} 
		\caption{(online color) {\it Probability curves for the dissipative state: Method A vs Method B.} Red and blue curves represent the results of non-unitary evolution obtained from Method A and Method B, respectively.}
		\label{fig:DissipationState}
	\end{figure}
	
	Fig. \ref{fig:ComparisonOpenOpen} compares the probability distributions for non-unitary evolution in an open system obtained from Method A and Method B. As can be seen in the Fig. \ref{fig:ComparisonOpenOpen}, the probability curves (red and blue) completely overlap before reaching the boundary (panel (a)). Once the boundary is reached, a slight difference begins to appear between the two curves (panel (b)). In the subsequent evolution, this difference does not amplify (panels (c)--(g)); instead, as the photon continues to dissipate into the external environment, the two curves tend to overlap again (panel (h)). It is noteworthy that throughout the evolution, the oscillations of the red and blue curves remain highly synchronized, with only a slight difference in amplitude. Specifically, the blue curve (Method B) is consistently lower than the red curve (Method A). This is because Method B includes more dissipation channels, allowing the photon to escape more quickly, and resulting in a lower probability of finding the photon in the two-dimensional plane.
	
	Fig. \ref{fig:DissipationState} compares the time evolution of the dissipative-state probability obtained from Method A and Method B. As shown in panel (a), in the early stages of evolution, before the photon reaches the boundary, both dissipative-state probability curves remain at zero. Once the photon reaches the boundary, the curves rise rapidly and no longer overlap (panel (b)), with the blue curve consistently above the red curve (panel (c)). This trend is opposite to that observed in Fig. \ref{fig:ComparisonOpenOpen}. Fig. \ref{fig:ComparisonOpenOpen} shows the probability distribution of the photon within the two-dimensional plane (i.e., the non-dissipative state), whereas Fig. \ref{fig:DissipationState} focuses on the dissipative state that arises once the photon has escaped the system.
	
	\begin{figure}
		\centering
		\includegraphics[width=1.\textwidth]{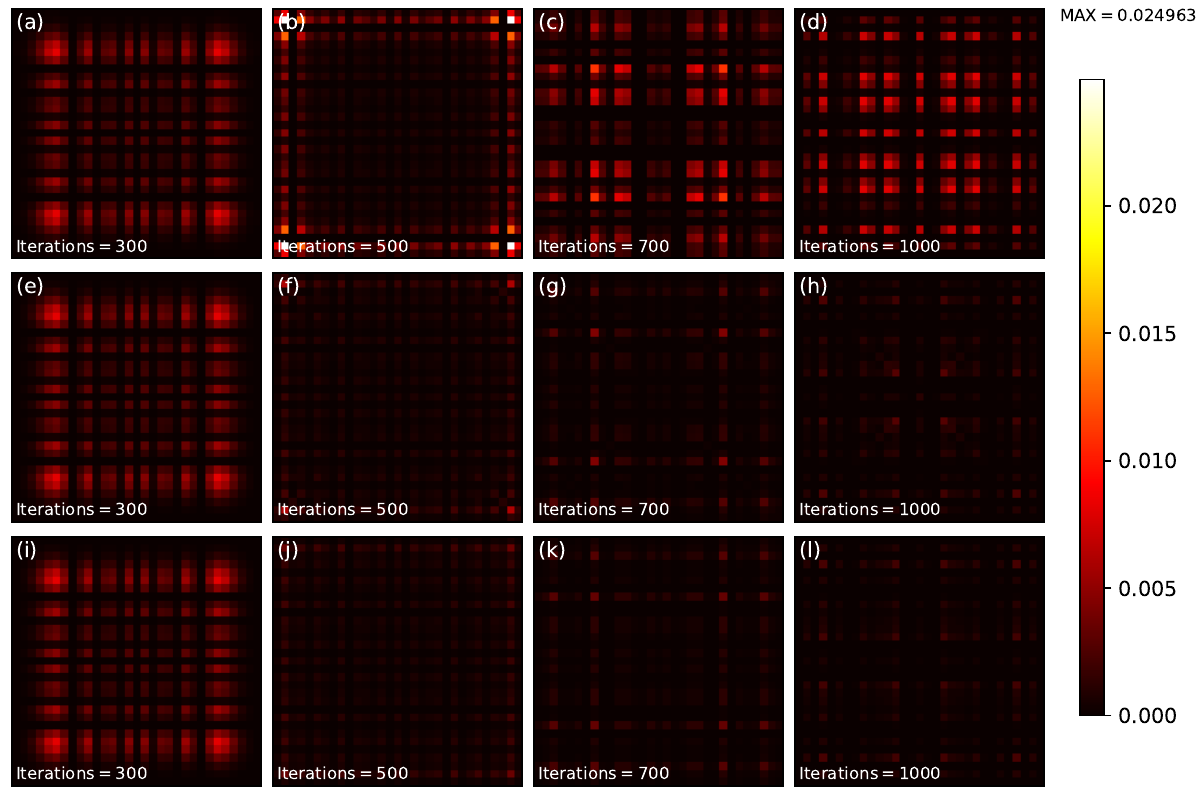} 
		\caption{(online color) {\it Probability distributions on the two-dimensional plane.} Comparison of three cases: closed system (first row), open system with Method A (second row), and open system with Method B (third row). Each row contains four panels corresponding to iteration steps 300, 500, 700, and 1000 (from left to right). The upper bound of the color bar is set to the global maximum probability of $0.024963$.}
		\label{fig:2DDistribution}
	\end{figure}

	Fig. \ref{fig:2DDistribution} illustrates the spatial probability distribution on the two-dimensional plane for three scenarios: a closed system (first row), an open system using Method A (second row), and an open system using Method B (third row). As can be seen from the figure, the spatial distributions in the three scenarios are identical before the probability distribution reaches the boundary. significant probability loss appears in the two open systems, and this loss increases rapidly with evolution. Notably, the difference in probability loss between the two open systems is very small, and their probability distribution patterns remain similar. This observation is consistent with the synchronized oscillations seen in Fig. \ref{fig:ComparisonOpenOpen}.

	\section{Conclusion and Outlook}
	\label{sec:Conclusion}
	
	This paper systematically studies the motion of a single photon in a two-dimensional plane, focusing on its dynamical evolution under closed and open boundary conditions. We construct the Hilbert space of the system using the standard second-quantization formalism (Method A) and a non-standard method (Method B), respectively. By eliminating redundant states, we significantly reduce the Hilbert space dimension, thereby improving the efficiency of numerical simulations.

	In the closed system, since the Hilbert spaces constructed by the two methods are equivalent, their unitary evolution results are identical. The probability distribution of the photon diffuses outward from the center of the plane and exhibits a significant bounce after reaching the boundary, reflecting the confining effect of the closed boundary.

	In the open system, the two methods are no longer equivalent. Method B, by incorporating more dissipation channels, provides a more accurate description of the photon escape process at the boundary. Numerical results show that the probability evolution curves obtained from the two methods completely overlap before reaching the boundary, and a slight difference appears once the boundary is reached. However, this difference does not amplify with evolution; instead, it tends to converge in the later stage when dissipation dominates. In particular, the dissipative-state probability obtained from Method B is consistently slightly higher than that from Method A, indicating that the photon escapes faster and has a lower residual probability in Method B.

	Through visualization of the probability distribution in two-dimensional space, we further validate the above conclusions. The three scenarios (closed system, open system with Method A, and open system with Method B) exhibit identical probability distributions before reaching the boundary. Afterward, significant probability loss occurs in the open systems, and this loss increases rapidly with evolution time. Although Method A and Method B differ in the number of dissipation channels, their probability distribution patterns are highly similar, demonstrating synchronized evolutionary behavior.

	This work can be further extended in the following directions. While the current model only considers the free motion of a single photon, future work could incorporate nonlinear interactions (such as the Kerr effect) to investigate multiphoton correlations and quantum phase transitions. Based on rectangular grids and simple boundary conditions, this work can be extended to more complex geometries (e.g., triangular or hexagonal lattices) or non-uniform boundaries to explore the influence of geometry on photon localization and escape.

	In summary, by comparing two Hilbert space construction methods, this paper reveals the motion of a single photon in a two-dimensional plane under closed and open boundaries, laying the foundation for future studies on photon transport and dissipation control in complex quantum networks.

	\begin{acknowledgments}
	The reported study was funded by China Scholarship Council, project number 202108090483.
	\end{acknowledgments}

	\bibliography{bibliography}

\end{document}